\documentclass[11pt]{article}
\usepackage{epsfig,a4}
\def\be{\begin{equation}}
\def\ee{\end{equation}}
\def\bea{\begin{eqnarray}}
\def\eea{\end{eqnarray}}
\def\nn{\nonumber}

\def\gev{\,{\rm GeV}}
\newcommand{\aem}{\alpha_{\rm em}}

\newcommand{\wf}{wave function}
\newcommand{\wfs}{wave functions}
\newcommand{\spa}{soft physics approach}
\renewcommand{\d}{\rm d}
\newcommand{\ibid}[1]{{\it ibid.}~#1}
\newcommand{\lsim}{\raisebox{-3pt}{$\,\stackrel{\textstyle <}{\sim}\,$}}

\begin{document}
    \setlength{\baselineskip}{2.6ex}

\title{Parton Distributions, Form Factors and Compton Scattering}

\author{P. Kroll \\
{\em Fachbereich Physik, Universit\"at Wuppertal}\\
{\em Gau\ss strasse 20, D-42097 Wuppertal, Germany} \\
\vspace{0.3cm}}
\date{}
\thispagestyle{empty}
\vspace*{1cm}
\begin{flushright}
WU B 99-24 \\
hep-ph/9910294\\[20mm]
\end{flushright}

\begin{center}
{\Large\bf Parton Distributions, Form Factors and Compton Scattering}\\[20mm]

{\Large P.\ Kroll}\\[10mm]
{\it Fachbereich Physik, Universit\"at Wuppertal}\\[2mm]
{\it Gau\ss strasse 20, D-42097 Wuppertal, Germany}\\[30mm]%
\end{center}

\begin{center}
Contribution to the Eighth International Symposium on Meson-Nucleon
Physics and the Structure of the Nucleon \\[5mm] 
Zuoz (August 1999)
\end{center}
\newpage
\setcounter{page}{1}

\maketitle

\begin{abstract}
\setlength{\baselineskip}{2.6ex}
The soft physics approach to form factors and Compton scattering
at moderately large momentum transfer is reviewed. It will be argued
that in that approach the Compton cross section is given by the
Klein-Nishina cross section multiplied by a factor describing the
structure of the proton in terms of two new form factors.
These form factors as well as the ordinary electromagnetic form
factors represent moments of skewed parton distributions.
\end{abstract}

\setlength{\baselineskip}{2.6ex}
QCD provides three valence Fock state contributions to proton form
factors, real (RCS) and virtual (VCS) Compton scattering  off protons at 
large momentum transfer: a soft overlap term with an active quark and two
spectators, the 
asymptotically dominant perturbative contribution where by means of
the exchange of two hard gluons the quarks are kept collinear
with respect to their parent protons and a
third contribution that is intermediate between the soft and the
perturbative contribution where only one hard gluon is exchanged and one of
the three quarks acts as a spectator. Both the soft and the intermediate terms
represent power corrections to the perturbative contribution. Higher
Fock state contributions are suppressed. The crucial
question is what is the relative strengths of the three
contributions at experimentally accessible values of momentum transfer,
i.e.\ at $-t$ of the order of 10 GeV$^2$? The pQCD followers assume
the dominance of the perturbative contribution and neglect the other two
contributions while the soft physics community
presumes the dominance of the overlap contribution. Which group is
right is not yet fully decided although comparison with the pion case \cite{kro96}
seems to favour a strong overlap contribution.  

\begin{figure}[h,b]
\begin{center}
   \psfig{file=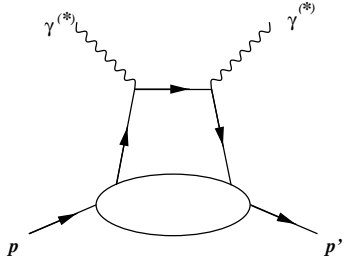,%
          bbllx=45pt,bblly=230pt,bburx=545pt,bbury=610pt,%
           width=4.5cm,clip=}
\end{center}
\caption{The handbag diagram for Compton scattering. The momenta of
         the incoming and outgoing protons (photons) are denoted by $p$ ($q$) 
         and $p'=p+\Delta$ ($q'=q-\Delta$), respectively. }
\label{fig:handbag}
\end{figure}

Let me turn now to the soft physics approach to Compton scattering.
For Mandelstam variables, $s$, $t$ and  $u$, that are large on a
hadronic scale the handbag diagram shown in Fig.\
\ref{fig:handbag} describes RCS and VCS. To see this it is of
advantage to choose a symmetric frame of reference where the plus and
minus light-cone components of $\Delta$ are zero.
This implies $t=-\Delta^2_\perp$ as well as a vanishing skewdness
parameter $\zeta=-\Delta^+/p^+$. To evaluate the skewed parton
distributions (SPD) appearing in the handbag diagram and defined in
\cite{mue98}, one may use a Fock state decomposition of the 
proton and sum over all possible spectator
configurations. The crucial assumption is then that the soft hadron
\wfs{} are dominated by virtualities in the range $|k_i^2|\lsim
\Lambda^2$, where $\Lambda$ is a hadronic scale of the order of 1 GeV,
and by intrinsic transverse parton momenta, $k_{\perp i}$, defined
with respect to their parent hadron's momentum, that satisfy $k_{\perp
i}^2/x_i\lsim \Lambda^2$. Under this assumption factorisation of the
Compton amplitude in a hard photon-parton subprocess amplitude and a
soft proton matrix element is achieved \cite{DFJK}. This proton matrix element is 
described by new form factors specific to Compton scattering.

As a consequence of this result the Compton amplitudes conserving the
proton helicity are given by
\be
{\cal M}_{\mu'+,\,\mu +} \,=\, \;2\pi\aem \left[{\cal
    H}_{\mu'+,\,\mu+}\,(R_V + R_A)
  \,+\, {\cal H}_{\mu'-,\,\mu-}\,(R_V - R_A) \right ]\,.
\label{final}
\ee
Proton helicity flip is neglected. $\mu$ and $\mu'$ are
the helicities of the incoming and outgoing photon in the
photon-proton cms, respectively. The photon-quark subprocess
amplitudes, ${\cal H}$, are calculated for massless quarks in lowest order QED.
The form factors in Eq.\ (\ref{final}), $R_V$ and
$R_A$, represent $1/x$-moments of SPDs at zero skewedness parameter.
$R_V$ is defined by
\begin{eqnarray} 
\sum_a e_a^2\, \int_0^1\, \frac{{\d} x}{x}\, p^+
   \int {{\d} z^-\over 2\pi}\, e^{i\, x p^+ z^-} 
     \langle p'|\,
     \overline\psi{}_{a}(0)\, \gamma^+\,\psi_{a}(z^-) - 
     \overline\psi{}_{a}(z^-)\, \gamma^+\,\psi_{a}(0) 
     \,| p \rangle  \nn \\[0.3em]
   = R_V(t)\, \bar{u}(p')\, \gamma^+ u(p)\, 
 + R_T(t)\, \frac{i}{2m}\bar{u}(p')
                          \sigma^{+\rho}\Delta_\rho u(p)\,,\hspace*{2cm}
\label{R-form-factors}
\eea
where the sum runs over quark flavours $a$ ($u$, $d$, \ldots), $e_a$
being the electric charge of quark $a$ in units of the positron
charge. $R_T$ being related to proton helicity flips, is neglected in
(\ref{final}). There is an analogous equation for the axial vector
proton matrix element, which defines the form factor $R_A$. 
Due to time reversal invariance the form factors $R_V$, $R_A$ etc.\
are real functions. 

As shown in \cite{DFJK} form factors can be represented as
generalized Drell-Yan light-cone \wf{} overlaps. Assuming a plausible
Gaussian  $k_{\perp i}$-depen\-dence of the soft Fock state \wfs{}, 
one can explicitly carry out the momentum integrations in
the Drell-Yan formula. For simplicity one may further assume a common
transverse size parameter, $\hat a$, for all Fock
states. This immediately allows one to sum over them, without
specifying the $x_i$-dependence of the \wf{}s. One then arrives at
\cite{DFJK,rad98a} 
\bea
F_1(t)&=& \sum_a\, e_a\, \int {\d} x\, 
         \exp{\left[\frac12 \hat a^2 t \frac{1-x}{x}\right]}    
                 \,\{ q_a(x) - \bar{q}_a(x) \} \,, \nn\\[0.5em]
R_V(t)&=& \sum_a\, e_a^2\, \int \frac {{\d} x}{x}\, 
         \exp{\left[\frac12 \hat a^2 t \frac{1-x}{x}\right]} 
                \,\{ q_a(x) + \bar{q}_a(x) \} \,,
\label{ffspd}
\eea
and the analogue for $R_A$ with $q_a+\bar{q}_a$ replaced by $\Delta
q_a + \Delta \bar{q}_a$. $q_a$ and $\Delta q_a$ are the usual
unpolarized and polarized parton distributions, respectively. 
The result for $F_1$ has been derived in Ref.\ \cite{bar} long time ago.

The only parameter appearing in (\ref{ffspd}) is the effective
transverse size parameter $\hat{a}$; it is known to be about 1
GeV{}$^{-1}$ with an uncertainty of about 20$\%$. Thus, this parameter
only allows some fine tuning of the results for the form factors.
Evaluating, for instance, the form factors from the parton
distributions derived by Gl\"uck et al. (GRV) \cite{GRV} with 
$\hat{a}=1\, GeV^{-1}$, one already finds good results.
Improvements are obtained by treating the lowest three Fock states
explicitly with specified $x$-dependencies \cite{DFJK}. 
Results for $t^2 F_1$ and $t^2 R_V$ obtained that way 
are displayed in Fig.\ \ref{fig:FFsoft}. Both the scaled
form factors, as well as $t^2 R_A$, exhibit broad maxima and,
hence, mimic dimensional counting rule behaviour in the $t$-range from 
about 5 to 15 GeV{}$^2$. The position, $t_0$, of the maximum of $t^2
F_i$, where $F_i$ is one of the soft form factors, is determined by
the solution of the implicit equation  
\begin{equation}
- t = 4 \hat a^{-2}\, \left\langle { \frac{1-x}{x}} \right\rangle^{-1}_{F_i,t}\,.
\label{maxpos}
\end{equation} 
The mean value $\langle \frac{1-x}{x} \rangle$ comes out
around $0.5$ at $t=t_0$, hence, $t_0\simeq 8 \hat{a}^2$. Since
both sides of Eq.\ (\ref{maxpos}) increase with $-t$
the maximum of the scaled form factor, $F_i$, is quite broad. 
For very large momentum transfer the form
factors turn gradually into the soft physics asymptotics $\sim
1/t^4$. This is the region where the perturbative contribution ($\sim
1/t^2$) takes the lead.  
\begin{figure}
\parbox{\textwidth}{\begin{center}
   \psfig{file=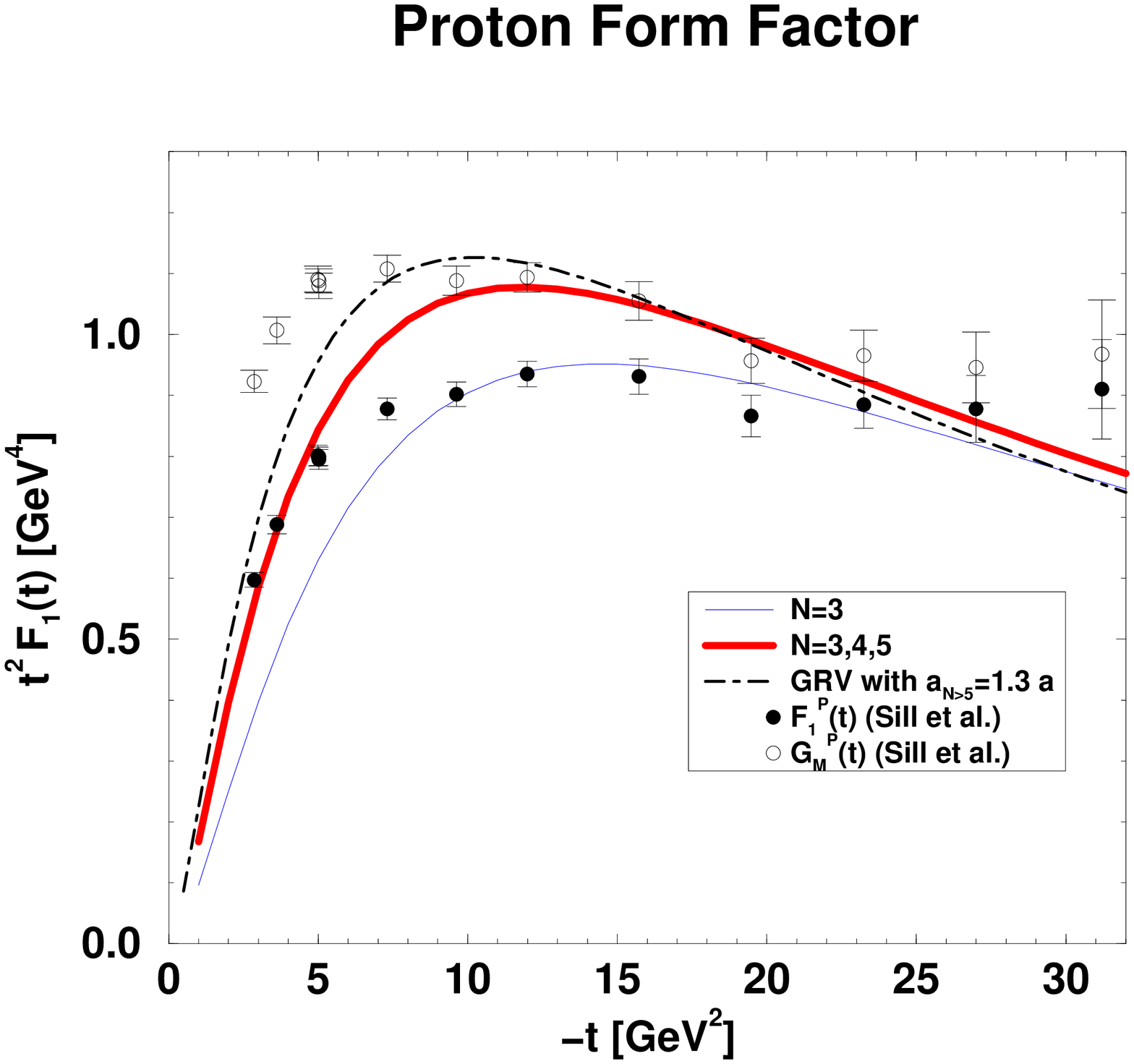, 
          bbllx=100pt,bblly=0pt,bburx=590pt,bbury=635pt,%
           width=5.3cm, angle=-90, clip=}\hspace{0.5cm}
      \psfig{file=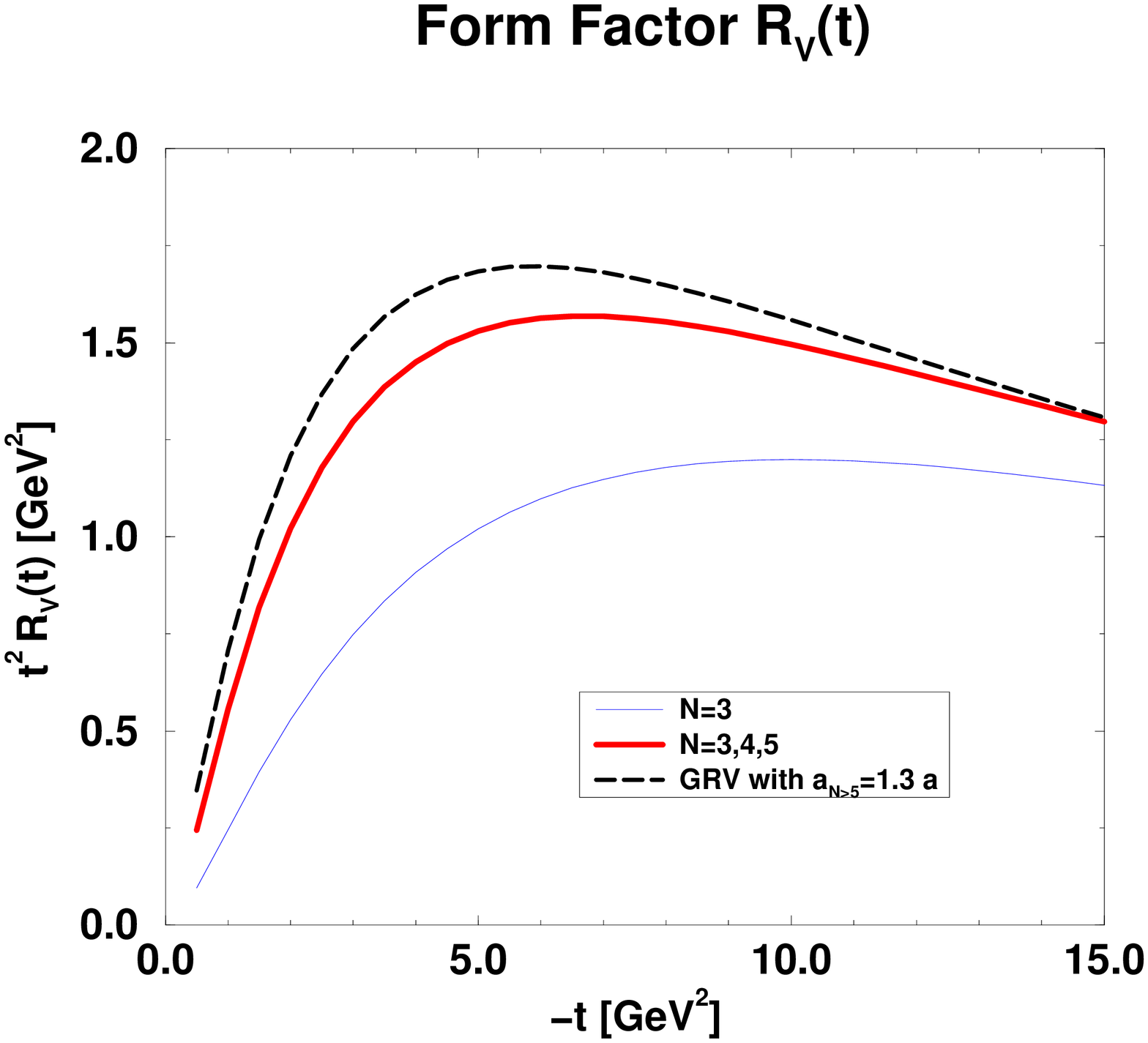, 
          bbllx=90pt,bblly=0pt,bburx=590pt,bbury=655pt,%
           width=5.3cm, angle=-90, clip=} 
\end{center}}
\caption{The Dirac (left) and the vector Compton (right) form factor
         of the proton as predicted by the soft physics approach 
         \protect\cite{DFJK,bol96}. Data are taken from
         \protect\cite{sil93}. The data on the magnetic form factor, $G_M$, are 
          shown in order to demonstrate the size of spin-flip effects.}
\label{fig:FFsoft}
\end{figure}


The amplitude (\ref{final}) leads to the RCS cross section  
\be
\frac{{\d} \sigma}{{\d} t} \,=\, \frac{{\d} \hat{\sigma}}{{\d} t}
                       \left [\, \frac{1}{2} (R_V^2(t) + R_A^2(t))
           -\, \frac{us}{s^2+u^2}\, (R_V^2(t)-R_A^2(t)) \,\right] \,.
\ee
It is given by the Klein-Nishina cross section 
\be
      \frac{{\d} \hat{\sigma}}{{\d} t}\, = \,\frac{2\pi\aem^2}{s^2}\; 
                         \frac{s^2+u^2}{-us}       
\ee
multiplied by a factor that describes the structure of the proton in
terms of two form factors. Evidently, if the form factors scale as
$1/t^2$, the Compton cross section would scale as $s^{-6}$ at fixed cm
scattering angle $\theta$. In view of the above discussion (see also
Fig.\ \ref{fig:FFsoft}) one therefore infers that approximate
dimensional counting rule behaviour holds in a limited range of energy.  
\begin{figure}
\parbox{\textwidth}{ \begin{center}
   \psfig{file=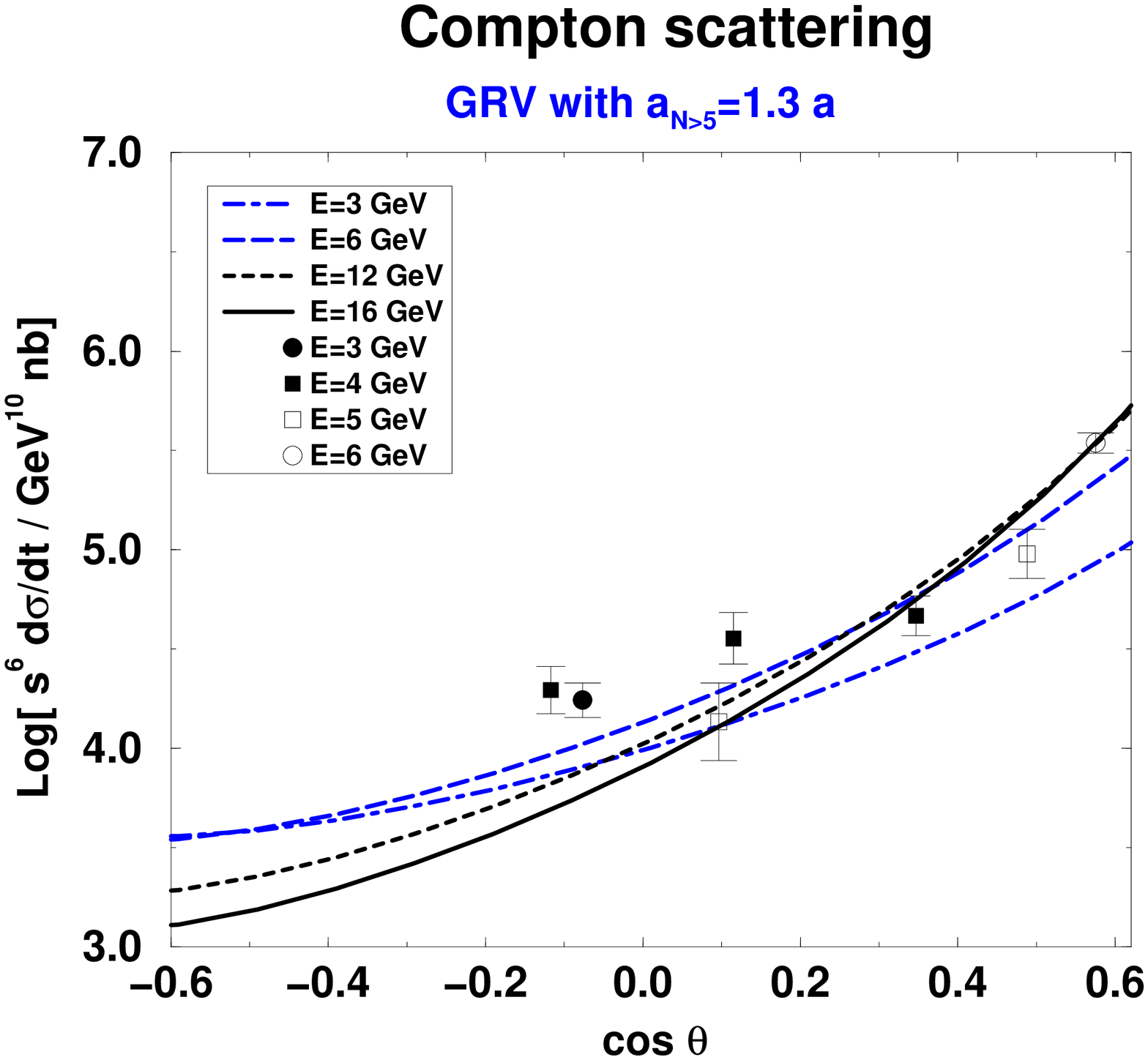, 
          bbllx=92pt,bblly=0pt,bburx=590pt,bbury=640pt,%
           width=5.5cm, angle=-90, clip=}\hspace{0.5cm}
           \psfig{file=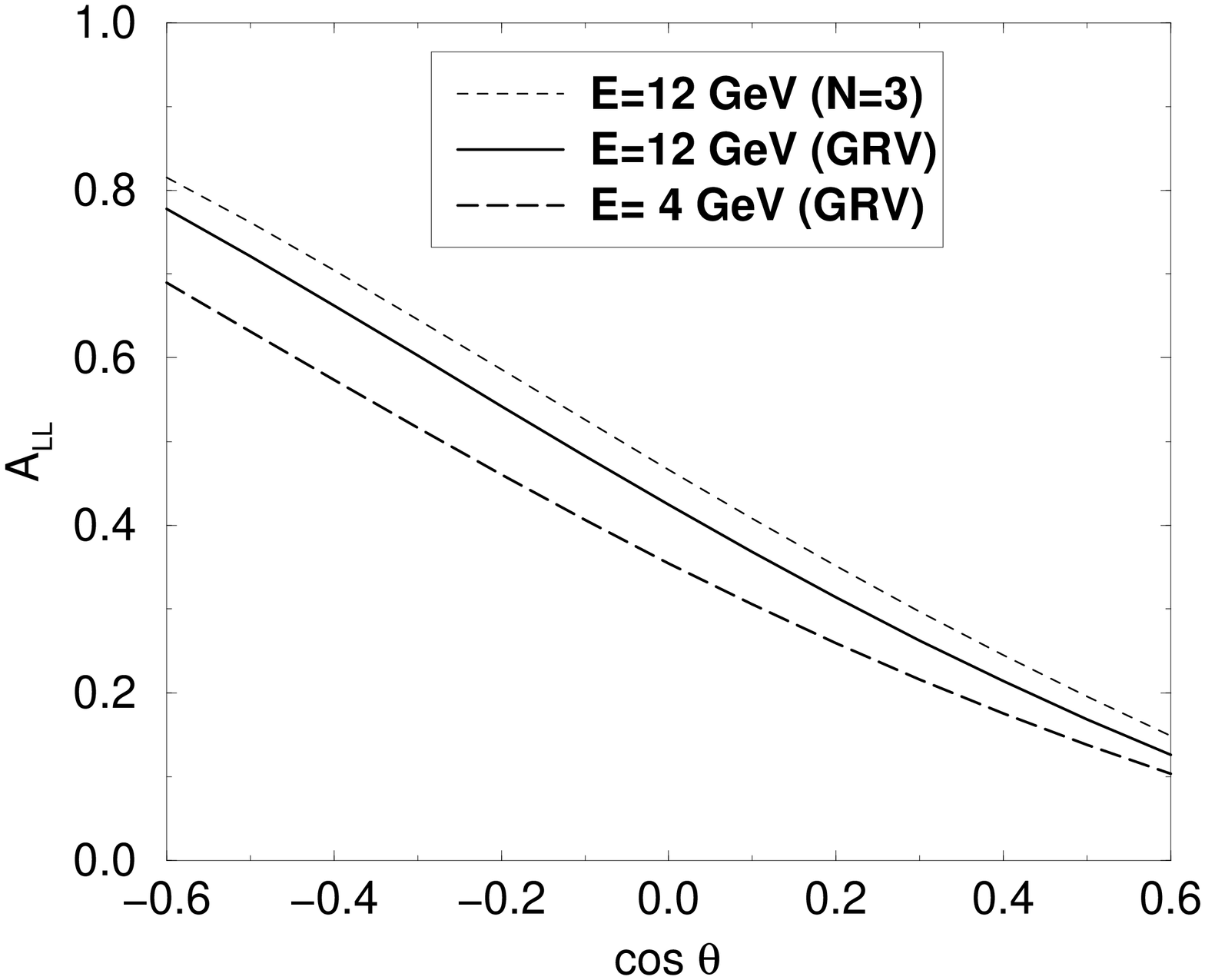, 
           width=5.5cm, angle=-90} 
\end{center}}
\caption{The Compton cross section, scaled by $s^6$, (left) 
    and the initial state helicity correlation $A_{\rm LL}$ (right)
    as predicted by the soft physics approach \protect\cite{DFJK}. Data taken
   from \protect\cite{shu79}.}
\label{fig:rcs}
\end{figure}
The magnitude of the Compton cross section is quite well predicted as
is revealed by comparison with the admittedly old data~\cite{shu79}
measured at rather low values of $s$, $-t$ and $-u$ (see Fig.\
\ref{fig:rcs}). 

The \spa{} also predicts characteristic spin dependencies of the
Compton process \cite{DFJK2}. Of particular interest is the initial state
helicity correlation
\be
A_{\rm LL}\, \frac{{\d} \sigma}{{\d} t} = \frac{2\pi\aem^2}{s^2} \;
  R_V(t) R_A(t) \left(\frac{u}{s} - \frac{s}{u}\right) \,.
\ee
Approximately, $A_{\rm LL}$ is given by the corresponding subprocess
helicity correlation $\hat{A}_{\rm LL}=(s^2-u^2)/(s^2+u^2)$ multiplied
by the dilution factor $R_A(t)/R_V(t)$. Thus, measurements of both the
cross section and the initial state helicity correlation allows one to
isolate the two form factors $R_V$ and $R_A$
experimentally~\cite{nat99}. In Fig.\ \ref{fig:rcs} predictions for
$A_{\rm LL}$ are shown. 

Other polarization observables as well as the VCS contribution to the
unpolarized $ep\to ep\gamma$ cross section have
also been predicted in \cite{DFJK2}. In addition to VCS the full 
$ep\to ep\gamma$ cross section receives substantial contributions from
the Bethe-Heitler process, in which the final state photon is radiated by the
electron. Dominance of the VCS contribution requires high energies,
small values of $|\cos{\theta}|$ and an out-of-plane experiment, i.e.\
an azimuthal angle larger than about $60^\circ$.
For VCS there are characteristic differences to the diquark model
\cite{kro96a}, the only other available study of VCS. In
the soft physics approach all amplitudes are real (if the
photon-parton subprocess is calculated to lowest order perturbation
theory \cite{DFJK,DFJK2}) while in the
diquark model there are perturbatively generated phase differences
among the VCS amplitudes. So, for instance, the beam asymmetry
for $ep\to ep\gamma$ 
\begin{equation}
A_{\rm L} = \frac{ {\d}\sigma (+) - {\d}\sigma (-)} 
                 { {\d}\sigma (+) + {\d}\sigma (-)} \,,
\end{equation}
where the labels $+$ and $-$ denote the lepton beam helicity,
is zero in the \spa in contrast to the diquark
model where a sizeable beam asymmetry is predicted.


In summary, the \spa{} leads to a simple representation of form factors and to
detailed predictions for RCS and VCS.
These predictions exhibit interesting features and
characteristic spin dependences with marked differences to other
approaches. Dimensional counting rule behaviour for form factors,
Compton scattering and perhaps for other exlusive observables is
mimicked in a limited range of momentum transfer.
This tells us that it is premature to infer the dominance of
perturbative physics from the observed scaling behaviour, see also \cite{ral}. 
The soft contributions although formally representing 
power corrections to the asymptotically leading perturbative
ones, seem to dominate form factors and Compton scattering
for momentum transfers around 10 \gev$^2$ (see the discussion in
\cite{bol96,ber95}). However, a
severe confrontation of this approach with accurate
large momentum transfer data on RCS and VCS is still pending.

\bibliographystyle{unsrt}

\end{document}